\documentclass[aps,prl,reprint,superscriptaddress,longbibliography]{revtex4-1}
\usepackage{amsfonts}
\usepackage{amsmath,amssymb}
\usepackage{units}
\usepackage{graphicx}

\begin{document}


\title{Fermi surface manipulation by external magnetic field demonstrated for a~prototypical ferromagnet}


\author{E.~{M\l y\'{n}czak}}
\email[]{e.mlynczak@fz-juelich.de}
\affiliation{Peter Gr\"unberg Institut PGI, Forschungszentrum J\"ulich and JARA- Fundamentals of Future Information Technologies, 52425 J\"ulich, Germany}
\affiliation{Faculty of Physics and Applied Computer Science, AGH University of Science and Technology, al. Mickiewicza 30, 30-059 Krak\'ow, Poland}

\author{M.~Eschbach}
\affiliation{Peter Gr\"unberg Institut PGI, Forschungszentrum J\"ulich and JARA- Fundamentals of Future Information Technologies, 52425 J\"ulich, Germany}

\author{S.~Borek}
\affiliation{Department Chemie, Ludwig-Maximilians-Universit\"at M\"unchen, Butenandtstrasse 5-13, 81377 M\"unchen, Germany}

\author{J.~Min\'ar}
\affiliation{Department Chemie, Ludwig-Maximilians-Universit\"at M\"unchen, Butenandtstrasse 5-13, 81377 M\"unchen, Germany}
\affiliation{New Technologies-Research Centre, University of West Bohemia, Univerzitni 8, 306 14 Pilsen, Czech Republic.}

\author{J.~Braun}
\affiliation{Department Chemie, Ludwig-Maximilians-Universit\"at M\"unchen, Butenandtstrasse 5-13, 81377 M\"unchen, Germany}

\author{I.~Aguilera}
\affiliation{Peter Gr\"unberg Institut PGI, Forschungszentrum J\"ulich and JARA- Fundamentals of Future Information Technologies, 52425 J\"ulich, Germany}

\author{G.~Bihlmayer}
\affiliation{Peter Gr\"unberg Institut PGI, Forschungszentrum J\"ulich and JARA- Fundamentals of Future Information Technologies, 52425 J\"ulich, Germany}

\author{S.~D\"oring}
\affiliation{Peter Gr\"unberg Institut PGI, Forschungszentrum J\"ulich and JARA- Fundamentals of Future Information Technologies, 52425 J\"ulich, Germany}

\author{M.~Gehlmann}
\affiliation{Peter Gr\"unberg Institut PGI, Forschungszentrum J\"ulich and JARA- Fundamentals of Future Information Technologies, 52425 J\"ulich, Germany}

\author{P.~Gospodari\v{c}}
\affiliation{Peter Gr\"unberg Institut PGI, Forschungszentrum J\"ulich and JARA- Fundamentals of Future Information Technologies, 52425 J\"ulich, Germany}

\author{S.~Suga}
\affiliation{Peter Gr\"unberg Institut PGI, Forschungszentrum J\"ulich and JARA- Fundamentals of Future Information Technologies, 52425 J\"ulich, Germany}
\affiliation{Institute of Scientific and Industrial Research, Osaka University, Ibaraki, Osaka 567-0047, Japan}

\author{L.~Plucinski}
\affiliation{Peter Gr\"unberg Institut PGI, Forschungszentrum J\"ulich and JARA- Fundamentals of Future Information Technologies, 52425 J\"ulich, Germany}

\author{S.~Bl\"ugel}
\affiliation{Peter Gr\"unberg Institut PGI, Forschungszentrum J\"ulich and JARA- Fundamentals of Future Information Technologies, 52425 J\"ulich, Germany}

\author{H.~Ebert}
\affiliation{Department Chemie, Ludwig-Maximilians-Universit\"at M\"unchen, Butenandtstrasse 5-13, 81377 M\"unchen, Germany}

\author{C.~M. Schneider}
\affiliation{Peter Gr\"unberg Institut PGI, Forschungszentrum J\"ulich and JARA- Fundamentals of Future Information Technologies, 52425 J\"ulich, Germany}


\date{\today}

\begin{abstract}
We consider the details of the near-surface electronic band structure of a prototypical ferromagnet, Fe(001). Using high resolution angle-resolved photoemission spectroscopy we demonstrate openings of the spin-orbit induced electronic band gaps near the Fermi level. The band gaps and thus the Fermi surface can be manipulated by changing the remanent magnetization direction. The effect is of the order of $\Delta$E = 100 meV and $\Delta \text {k} = 0.1\,\text{\AA}^{-1}$. We show that the observed dispersions are dominated by the bulk band structure. First-principles calculations and one-step photoemission calculations suggest that the effect is related to changes in the electronic ground state, rather than caused by the photoemission process itself. The symmetry of the effect indicates that the observed electronic bulk states are influenced by the presence of the surface, which might be understood as related to a Rashba-type effect. By pinpointing the regions in the electronic band structure where the switchable band gaps occur, we demonstrate the significance of spin-orbit interaction even for elements as light as \textit{3d} ferromagnets. 
\end{abstract}


\maketitle

\section{I. INTRODUCTION}
The electronic band structure near the Fermi level determines numerous vital properties of metallic materials, being responsible for their thermal, magnetic and electronic transport behavior. In case of the metallic ferromagnets, the electronic band structure is split into the minority and majority spin states, as a result of the exchange interaction which rules the relative arrangement of the spins. What binds the spin direction to the orbital degrees of freedom is a relatively weak coupling, the spin-orbit interaction (SOI). The influence of the SOI on the electronic band structure of a ferromagnet is very subtle. It causes mixing of the spin character and a magnetization-dependent opening of minute energy gaps ($\sim ∼100$ meV), but only in specific points of the reciprocal space. The consequences of these delicate modifications are, however, tremendous and especially striking in the field of magnetism. One of them is the occurrence of the magnetocrystalline anisotropy (MCA), which ties the magnetization vector to specific axes of a crystal \cite{vanderLaan_MCA}. Moreover, the widely used experimental technique of the x-ray magnetic linear dichroism provides information about the magnetic moment direction, because of the dependence of the unoccupied electronic band structure on the magnetization direction \cite{vanderLaan1999}. The SOI-related modifications of the electronic structure near the Fermi level are also the basis for a new generation of spintronic devices. In 2002, it was experimentally shown by scanning tunneling microscopy, that tunneling current depends on the direction of sample magnetization, even when measured with the nonmagnetic tip \cite{Bode2002}. This effect, occuring due to the difference of the densities of states near the Fermi level for two orthogonal in-plane magnetization directions of the electrode was later termed tunneling anisotropic magnetoresistance (TAMR) \cite{Gould2004}. It was proposed \cite{Gould2004}, that thanks to TAMR it is possible to realize a spin-valve function using only one ferromagnetic (FM) electrode. A similar phenomenon, called ballistic anisotropic magnetoresistance (BAMR), was reported for structures where ballistic conductance takes place, such as nanowires \cite{Velev2005}. BAMR occurs, because the conductance is directly related to the number of open conducting channels, which might be different along and perpendicular to the nanowire \cite{Velev2005}. Another spintronic effect, for which SOI plays a major role is spin-orbit torque (SOT), recently discovered for heterostructures where a large exchange coupling and high SOI coexist \cite{Chernyshov2009,MihaiMiron2010}. Thanks to SOT it is possible to switch the magnetization direction using a charge current that flows in-plane of the heterostructure, which is very promising for spintronic applications \cite{Fukami2014}. 

Here we report on the direct experimental observation of the magnetization-dependent opening of the SOI-induced energy gaps near the Fermi level. As a subject of this study, we chose Fe(001) thin films grown epitaxially on an Au(001) single crystal, which can be considered as a prototypical magnetic system. Fe grown on Au(001) is characterized by a very low lattice mismatch (0.6$\%$) which results in minute strain experienced by the Fe film. Various magnetic and electronic phenomena such as ferromagnetism in the monolayer (ML) regime \cite{Bader1987}, existence of quantum well states \cite{Himpsel1991} or the thickness-driven spin reorientation transition \cite{Wilgocka2010} have been studied in this system in the past. 
We analyzed the electronic structure of the Fe(001)/Au(001) system using angle-resolved photoemission spectroscopy (ARPES) for four different remanent in-plane magnetization directions of the FM film, thus exploring the electronic principles of the SOI-related effects. Our experiments revealed distinct changes in the position of electronic bands near the $\overline{\text{X}}$ point of the surface Brillouin zone (SBZ) in response to the change of the magnetization direction. To interpret the experimental results in terms of the bulk electronic structure we performed calculations using the Green’s function formalism within the \textit{GW} scheme. For the discussion of the electronic structure of the Fe(001) surface we employed slab calculations based on the generalized gradient approximation (GGA). A theoretical treatment of the entire photoemission process was addressed using state-of-the-art one-step model photoemission calculations.

\section{II. EXPERIMENT AND CALCULATION}
\subsection{A. Experiment}
The 100~ML Fe films (1$\,$ML = 1.43$\,\text{\AA}$) investigated in this study were deposited by molecular beam epitaxy onto the Au(001) single crystal surface. The Au(001) template was prepared by repeated cycles of Ar sputtering and subsequent annealing at 500$^\circ$C for 10 min until the well-known surface reconstruction \cite{Hammer2014} was clearly visible using low energy electron diffraction (LEED). During the Fe deposition and ARPES measurement with the Neon emission line, the Au(001) substrate was kept at 50 K to prevent condensation of the Neon atoms on the surface. After deposition, the Fe films were shortly heated up to 300$^\circ$C. 
Before each ARPES measurement, the samples were remanently magnetized. The external magnetic field was applied to the thin film sample by an oriented permanent magnet ($\sim$20 mT), which was approached to the sample surface (as close as 2-3 cm) in an air-well set on a linear feedthrough. The procedure of the magnetization was similar to the one used in Ref. \cite{Plucinski2010}. For the thickness range used in this study, the easy magnetization directions of Fe(001) are all $<$100$>$ directions lying in the (001) plane \cite{Wilgocka2010}, and due to the low coercivity (${\text{H}_{\text{C}}}$= 1-2 mT), a remanent magnetization is easily realized. Stray fields caused by the remanent magnetization of the Fe(001) film do not distort the trajectories of emitted photoelectrons due to the very small Fe volume. ARPES measurements were performed for samples magnetized along each of the in-plane easy magnetization directions of Fe(001), \textit{i.e.}, [100], [-100], [010], and [0-10], which will be referred to as UP, DOWN, RIGHT and LEFT, respectively. 
\\The ARPES experiments were performed with a laboratory-based MBS A-1 electron analyzer using unpolarized non-monochromatized Neon emission from a SPECS UVS-300 discharge lamp with focusing capillary. All the ARPES spectra that will be discussed in this study were collected in the energy region close to the Fermi level (within a binding energy range of 200 meV), utilizing the higher-energy Neon emission line of $h\nu$~=~16.85 eV. Figure 1 shows a sketch of the experimental geometry. The analyzer was set to an energy resolution of 10 meV for all the presented spectra. The long axis of the analyzer entrance slit is parallel to the \textit{x}-axis corresponding to the detected electron emission plane. Further, our reference frame defines the measured electron wave vectors $k_{\parallel x} $ and $k_{\parallel y}$ as parallel and perpendicular to the entrance slit, respectively. When electrons are detected along the sample normal ($\Theta = 0°$), the light impinges under a grazing angle of $\phi\,=\,15^\circ$ with respect to the sample surface which lies in the \textit{xy}-plane of the laboratory reference frame. To measure the Fermi surface, the sample was rotated around the \textit{x}-axis, as the change in angle $\Theta$ corresponds to change in $k_{\parallel y}$, according to: $k_{\parallel y}=\sqrt{\frac{2m}{\hbar^{2}} E_{kin}} \sin{\Theta}$, where $E_{kin}$ represents the kinetic energy of photoemitted electrons. Therefore, the light incidence angle ($\phi = 15^\circ+\Theta$) changed during scanning of the angle $\Theta$, reaching $53^\circ$ at the $\overline{\text{X}}$ point. 

 \begin{figure}
 \includegraphics[width=8cm]{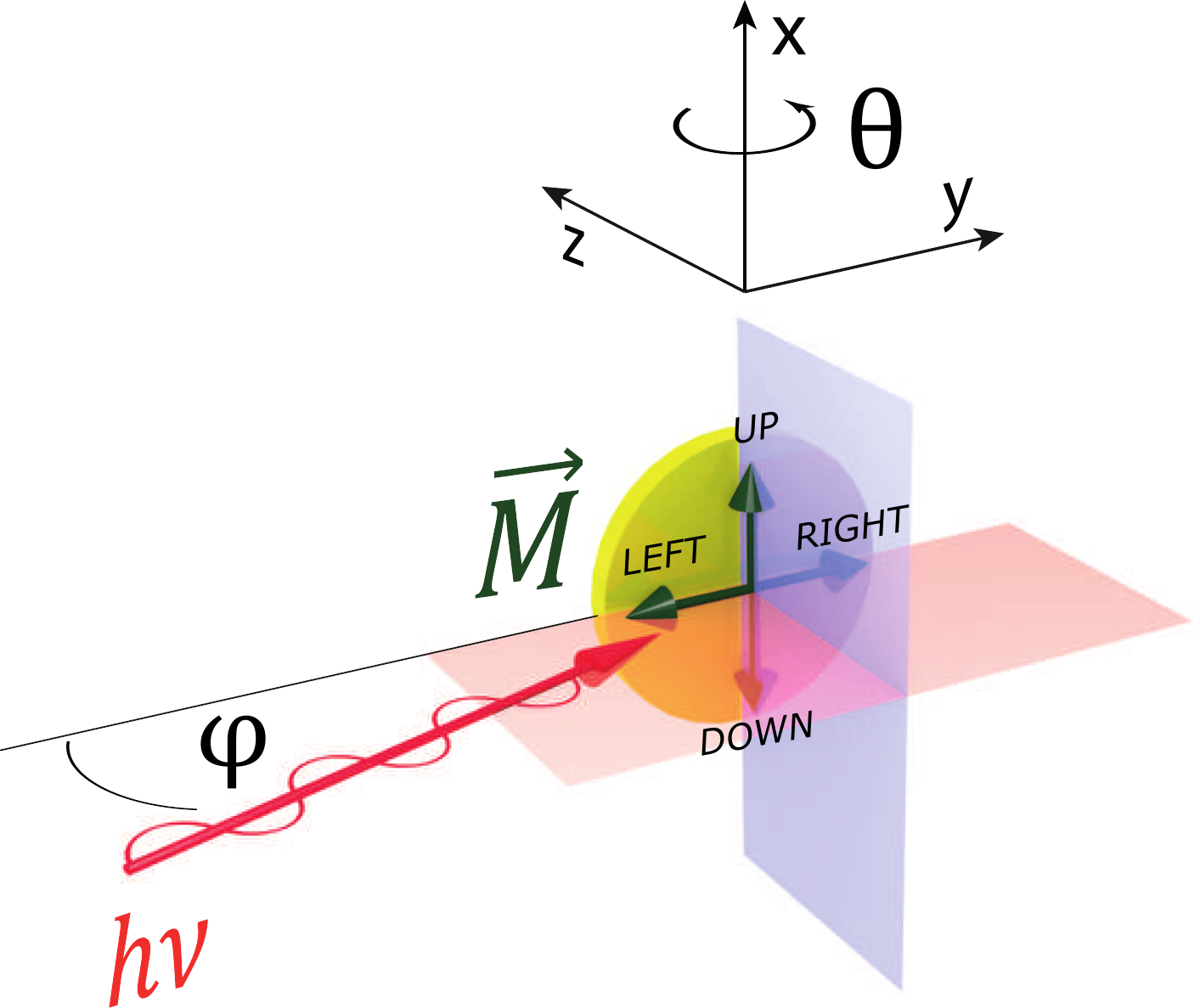}%
 \caption{Schematic representation of the experimental ARPES geometry. Light incidence and electron emission planes are indicated by a red (horizontal) and a blue (vertical) plane, respectively. Black arrows represent magnetization directions referred to as UP, DOWN, LEFT and RIGHT.\label{Fig1}}
 \end{figure}

\subsection{B. Calculation}
In order to obtain the theoretical bulk band structure of bcc Fe, many-body calculations in the framework of the \textit{GW} approximation were carried out within the all-electron full-potential linearized augmented-plane-wave (FLAPW) formalism as implemented in the SPEX code \cite{Friedrich2010}. 
For simulating the surfaces, we employed density functional theory (DFT) in the generalized gradient approximation (GGA) in the form of Perdew \textit{et al.} \cite{Perdew1996}. We used 27 layers of Fe(001), for the Au-covered surfaces symmetrically terminated with a p($1\times1$) Au ML, and relaxed the outermost four layers using the film version of the FLEUR code \cite{FLAPW}. Spin-orbit coupling was included self-consistently in the calculations. For the assessment of the Fermi-surfaces a $39\times39$ \textit{k}-point grid was used.
\\To theoretically analyze the photoemission process, one-step model photoemission calculations were performed. These calculations can be subdivided into two main parts. In the first part one has to determine the electronic structure of the system. Therefore, we set up a semi-infinite system which simulates the Fe(001) surface. For the calculation of the ground-state properties we used a fully relativistic multiple scattering method in the framework of density functional theory (Korringa-Kohn-Rostoker, KKR) \cite{Ebert2012}. In the second part the fully relativistic one-step model of photoemission was applied. The approach goes back to the developments worked out by Pendry and coworkers \cite{Pendry1974,Pendry1976,Hopkinson1980}. We calculated the elastic part of the photocurrent and neglected the interaction of the outgoing photoelectron with the rest of the electronic system (sudden approximation). For the calculation of the matrix elements which define the transition probability of the photoelectron one has to consider the initial and final state wave functions. The final state has been constructed using the theory of spin-polarized low energy electron diffraction (SPLEED). In this framework the final state is represented by a so-called time-reversed SPLEED state \cite{Braun1996,Braun2001}. Using a parametrized and energy-dependent inner potential we address the many-body interactions. We therefore corrected the elastic part of the photocurrent for inelastic interactions phenomenologically. For the escape of the photoelectron into vacuum one has to consider a surface barrier for which we used the parametrization of Rundgren and Malmstr\"om \cite{Malmstroem1980}. The barrier can be treated straightforward as additional surface layer and accounts for surface contributions to the photocurrent. Furthermore we can investigate the various possible transitions separately by the suppression of predetermined initial states which gives the possibility to analyze the main contribution to the transitions according to the dipole selection rules. Because the light source used in the experiment is unpolarized, we consider a $50\%$ mixture of the results obtained for the calculations for \textit{p}- and \textit{s}-polarized light.

\section{III. RESULTS AND DISCUSSION}
\subsection{A. Spin-orbit coupling signatures in the bulk electronic band structure of bcc Fe}
Figure 2 (a) shows the sketch of the bcc bulk Brillouin zone (BBZ) and the (001) surface Brillouin zone (SBZ), which we will refer to frequently when discussing electronic band dispersions. An arrow indicates the magnetization direction. Note, that the $\overline{\text{X}}$ point is the projection along $k_\parallel = 0.5|\Gamma$ - $\text{H}|$. Band dispersions found for the $\Gamma$ - H ($\Delta$) line of the bulk Brillouin zone using the \textit{GW} method are presented in Fig. 2 (b). Two directions are distinguished: parallel and perpendicular to the magnetization ($\vec{\text{M}}$) (plotted for negative and positive \textit{k} values, respectively). The color code refers to the spin character of each band; blue (red) marks minority (majority) spin states. Due to the introduction of spin-orbit interaction, electronic states of opposite spin become significantly mixed in the vicinity of the points where the bands would cross if no spin-orbit interaction was present. As a result, not only purely minority and majority states exist, but also states with mixed spin character. The bands are marked according to the single group representation, which is frequently used in the literature, especially in the discussions of the tunneling effect in the Fe/MgO/Fe magnetic tunnel junctions \cite{Butler2001}. The small Greek letters ($\alpha, \beta, \gamma, \delta$) correspond to the labeling of the Fermi sheets used later in this article to identify the experimentally observed spectral signatures.

\begin{figure}
 \includegraphics[width=8cm]{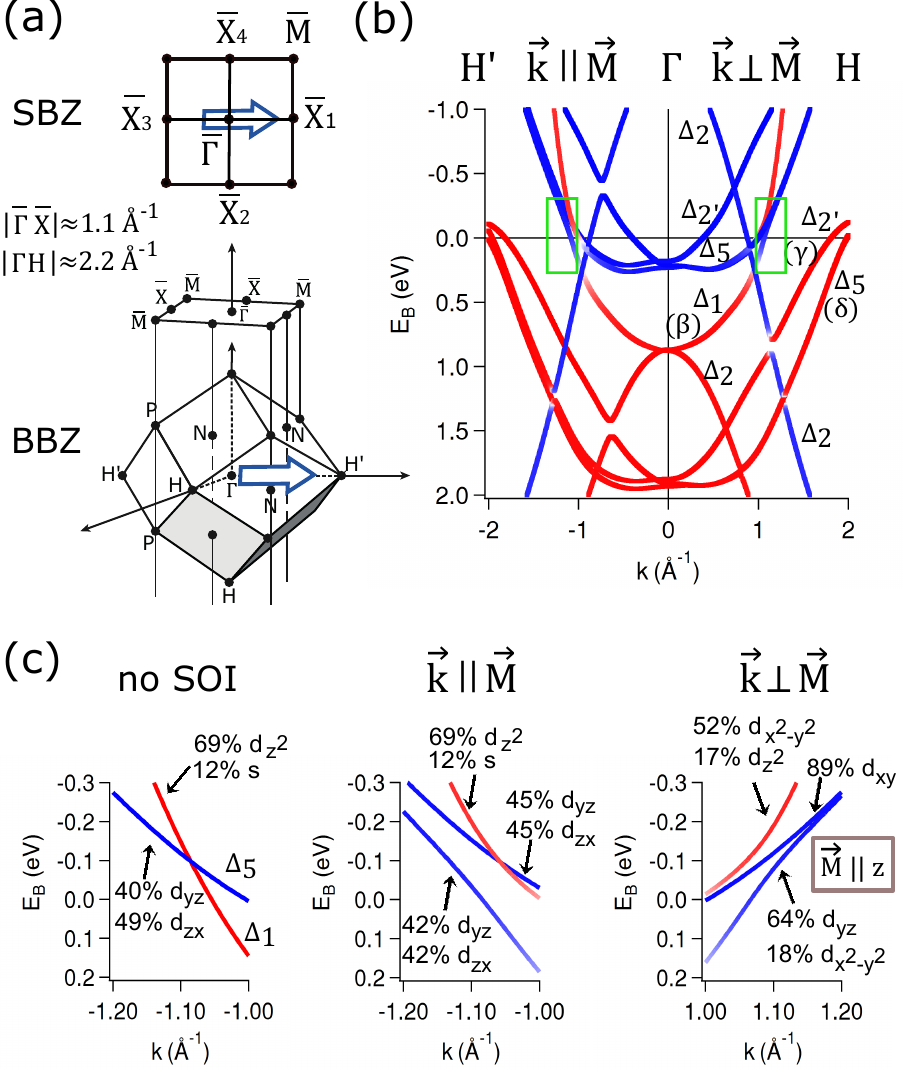}
 \caption{
 	(a) Bulk and (001) surface Brillouin zones (BBZ and SBZ) of bcc Fe. High symmetry points H and H$^\prime$ as well as $\overline{\text{X}}_1$  to $\overline{\text{X}}_4$ are defined with respect to the magnetization direction depicted by an arrow. (b) Relativistic bulk band structure of bcc Fe, calculated along the $\Gamma$ - H (positive \textit{k}) and $\Gamma$-H$^\prime$ (negative \textit{k}) using the \textit{GW} method. Green rectangles mark regions enlarged in (c). (c) Place of interest near $k_\parallel$ = 0.5$|\Gamma$ - $\text{H}|$, close to the Fermi level (shown by green rectangles in (b)) where three bands cross in the magnetization-dependent manner. Left panel shows the non-relativistic situation (no SOI), in the center panel the \textit{k}-vectors are parallel to $\vec{\text{M}}$, and the right plot shows \textit{k} vectors perpendicular to $\vec{\text{M}}$. Blue (red) color marks predominantly minority (majority) spin. The magnetization ($\vec{\text{M}}$) is parallel to the \textit{z}-axis.
 }	\label{Fig2}
 \end{figure}

We observe considerable differences between the band structures for the direction along the magnetization $\vec{\text{M}}$ and perpendicular to the magnetization (Fig. 2 (b)).  For example, the crossing of the minority spin $\Delta_2$ and $\Delta_2$$'$ bands right above the Fermi level develops a gap for the direction parallel to the magnetization. Such an avoided crossing is also observed for the minority $\Delta_2$ and majority $\Delta_2$$'$ bands that cross near $E_B$ = 1eV for the \textit{k} vectors along the magnetization, but show a gap opening for the perpendicular direction. As expected, crossing points for which hybridization of the bands occurs and the spin-orbit energy gaps open depend not only on the spatial part of the respective wavefunctions, but also on the direction of the spin within each band. The size of the spin-orbit gaps reaches $\Delta E_{SO}\sim$ 100 meV.
 
The electronic band structure near the Fermi level is of particular interest due to its influence onto the magnetocrystalline anisotropy and the electronic transport. Therefore, we identified the places of interest in the bulk Brillouin zone along the $\Delta$ line that showed magnetization-dependent openings of spin-orbit gaps. These regions are marked by green rectangles in Fig. 2 (b) and shown in detail in Fig. 2 (c). For the determination of the orbital symmetries, we define the magnetization $\vec{\text{M}}$ as being parallel to the \textit{z}-axis. The predominant orbital characters of each band, which govern the occurrence of the SOI-induced hybridization gaps, are listed as follows: on the left hand side of Fig. 2 (c) we plot the dispersion of the $\Delta_1$ band (predominantly d$_{z^2}$ orbital symmetry) and the doubly degenerated $\Delta_5$ band (d$_{yz}$ + d$_{zx}$ orbital symmetry) calculated without spin-orbit interaction. In the relativistic case, \textit{i.e.}, when the spin-orbit interaction is included in the calculation (Fig. 2 (c) middle and right), the $\Delta_5$ band splits and hybridizes with the $\Delta_1$ band, as $<\Psi (d_{z^2}\,\uparrow)|H_{SO}|\Psi (d_{yz(zx)}\,\downarrow)> \neq 0$. For the \textit{k}-vectors parallel to $\vec{\text{M}}$, the bands of d$_{z^2}$ and d$_{yz}$ (d$_{zx}$) symmetry can still cross. However, for the \textit{k}-vectors perpendicular to $\vec{\text{M}}$, all three bands avoid crossing each other. Therefore, we expect an experimentally observable difference between these two directions in \textit{k}-space in the vicinity of the $k_\parallel$ = 1.0 - 1.2$\,\text{\AA}^{-1}$, \textit{i.e.}, near the $\overline{\text{X}}$ point of the SBZ. 

\subsection{B. Band structure of Fe/Au(001): experiment and theory}

ARPES spectra were recorded for a series of polar angles $\Theta$ spanning the range of the SBZ between the $\overline{\Gamma} (k_y = 0\,\text{\AA}^{-1}$) and $\overline{\text{X}}$ points ($k_y\sim$ 1.1 $\,\text{\AA}^{-1}$) (see Fig. 1 and Fig. 2 (a) for the sketch of the surface Brillouin zone). The obtained data set measured for the sample magnetized RIGHT is shown in Fig. 3 as a band map along the $\overline{\Gamma}$ - $\overline{\text{X}}$ line (a) and the corresponding constant energy cut at the Fermi level ($E_F$($k_x$,$k_y$), later referred to as the Fermi surface) (b). The observed electronic bands are sharpest near the Fermi level, but become diffuse for larger binding energies of the order of 100 meV, possibly due to the influence of the electron correlation effects in the system \cite{Katsnelson1999}. For binding energies larger than $E_B$ = 180 meV a spectral intensity from the second Neon emission line ($h\nu$ = 16.67 eV) is visible.

\begin{figure*}
	\includegraphics{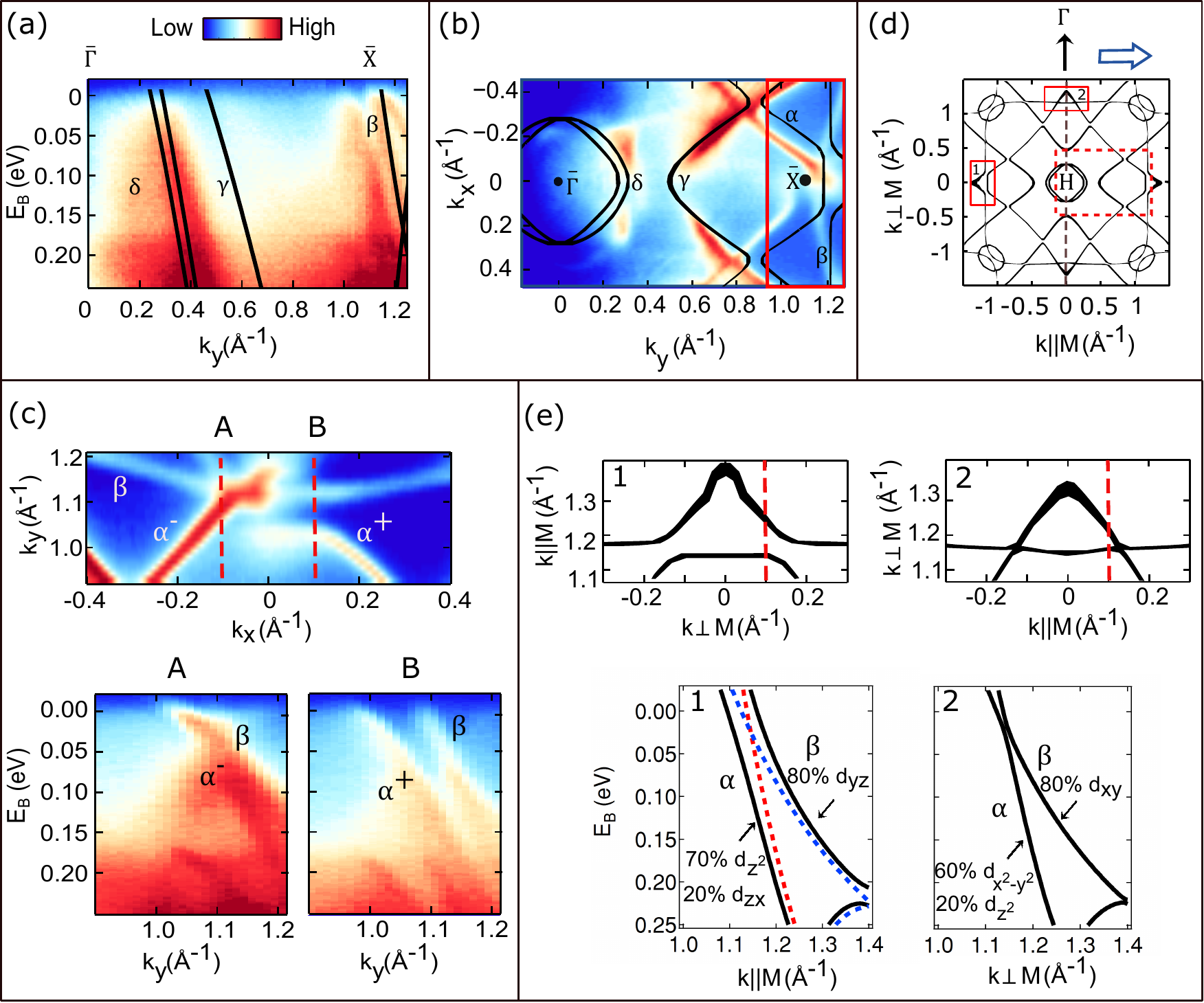}
	\caption{Fe(001) electronic band structure measured for the magnetization RIGHT. (a) ARPES spectra obtained for the binding energies close to the Fermi level, along $\overline{\Gamma}$ - $\overline{\text{X}}$ direction (b) Fermi surface in the vicinity of the $\overline{\Gamma}$ - $\overline{\text{X}}$ direction. The red rectangle on the right marks the region where detailed spectra for other magnetization directions were obtained (Fig. 5(a)). Black lines in (a) and (b) are superimposed results of the \textit{GW} calculations of the bulk electronic structure of Fe for $k_{\perp} \sim 2.2\, \text{\AA}^{-1}$. Calculated dispersions in (a) correspond to the  H$^\prime$-$\Gamma$ direction depicted in Fig. 2 (b). (c) Zoom of the Fermi surface measured in the region of the red rectangle marked in (b) (90$^\circ$ rotated) together with the dispersions for $|k_x| = 0.1\, \text{\AA}^{-1}$ (along two dashed red lines) marked as \textit{A} and \textit{B}. (d) \textit{GW} calculations of the bulk electronic structure of Fe for $k_{\perp} \sim 2.2\, \text{\AA}^{-1}$. The red dashed rectangle marks the \textit{k}-space region measured in the experiment. (e) Zooms within the area of the two red rectangles marked in (d) together with the theoretical dispersions along the red dashed lines. Dashed lines in the bottom left panel are the bands calculated without the introduction of SOI: blue (red) indicates minority (majority) spin.
	\label{Fig3}}
\end{figure*}

Figures 3 (a) and (b) show that the experimentally observed electronic bands were found to match reasonably well the calculations of the \textit{GW} bulk electronic structure of bcc Fe for the experimental Fermi level corresponding to $E_B$ = 0.1 eV in the theoretical result and for $k_{\perp} \sim 2.2\, \text{\AA}^{-1}$, \textit{i.e.}, near the H point, in agreement with the free electron final state model (see APPENDIX \textit{C} for the details). The calculated dispersions presented in Fig. 3 (a) correspond to the H$^\prime$-$\Gamma$ direction depicted in Fig. 2 (b). The Fermi surface sheets and the corresponding bands are marked with small Greek letters. We can identify characteristic shapes of the $\alpha$ and $\gamma$ sheets (although $\alpha$ sheet cannot be traced along the $\overline{\Gamma}$ - $\overline{\text{X}}$ line in Fig. 3 (a), it is very clearly observed for non-zero $k_x$). Another prominent feature that we find close to the $\overline{\text{X}}$ point is the $\beta$ sheet that in three dimensions ($k_x$, $k_y$, $k_z$) exhibits a cube-like shape. In the ARPES spectra we also observe a significant spectral contribution near $k_y\sim$ 0.2$\,\text{\AA}^{-1}$, which can be attributed to the Fermi sheet $\delta$ even though the spin-orbit splitting of this state cannot be resolved. We will show below in Fig. 5 that the experimental spectral intensity in this region of \textit{k}-space is well reproduced by one-step model photoemission calculations. 
 
The \textit{k}-space region measured in the experiment near the $\overline{\text{X}}$ point is magnified in Fig. 3 (c). In addition, band dispersions for wave vectors with $|k_x| =\,0.1\,\text{\AA}^{-1}$ are shown in the lower left and right panels \textit{A} and \textit{B} (cuts along the lines \textit{A} and \textit{B} marked in the upper panels). Clearly, the intensity distribution is much different between states with negative and positive $k_x$ values: two bands located closer (in cut \textit{A}) or clearly separated from each other (cut \textit{B}) are experimentally observed. To describe this effect, we term one branch of the $\alpha$ sheet $\alpha^+$ and the other $\alpha^-$ (branches found for positive and negative $k_x$ values, respectively). The question arises, whether the difference between the dispersions of the bands $\alpha^+$ and $\alpha^-$ can be explained by the occurrence of the SOI-related band gaps in the electronic structure of bulk bcc Fe? To answer this question, we need to analyze the total symmetry of the bulk electronic structure of Fe (Fig. 3 (d)) as well as the details of the band dispersions near $k_\perp = 0.5|\text{H}$ - $\Gamma|\,(\sim  1.1\, \text{\AA}^{-1}$, \textit{i.e.}, close to the $\overline{\text{X}}$ point) (Fig. 3 (e)). 

Results of the \textit{GW} calculation for the constant energy cut through the bulk electronic structure at the H point ($k_{\perp} \sim 2.2\, \text{\AA}^{-1}$) within the area of the entire surface Brillouin zone are presented in Fig. 3 (d). The magnetization direction is marked by an arrow. We can clearly see here the effect that was introduced in Fig. 2 (b), namely, that the \textit{k}-space directions along and perpendicular to the magnetization are not equivalent. Indeed, in the ferromagnetic state, the symmetry of the electronic structure is lower than the full octahedral symmetry of the atomic structure of the Fe bulk crystal (described by the point group $O_h$ \cite{Crystallography}). A single mirror plane that can be found in this picture is marked with the dashed vertical line. Additionally, a 4-fold rotation axis along the magnetization vector exists, leading to the symmetry group $D_{4h}$. The small red rectangles indicated by the numerals 1 and 2 mark the region of the \textit{k}-space close to the $k_\perp = 0.5|\text{H}$ - $\Gamma|$ (near the $\overline{\text{X}}$ point) where the magnetization-dependent opening of the SOI-related gaps occurs. The Fermi sheets within the areas 1 and 2 are magnified in Fig. 3 (e). This magnetization-related modification of the Fermi sheets was introduced in Fig. 2 (c), where the corresponding dispersions along the $\Gamma$ - H line were presented. Here (Fig. 3 (e)), we additionally show the dispersions along the red dashed lines ($0.1\, \text{\AA}^{-1}$ away from the high-symmetry line $\Gamma$ - H). The two bands show an avoided crossing for both $ k\parallel\vec{\text{M}}$ and $k\perp\vec{\text{M}}$ which can be seen by the comparison with the dispersions calculated without the spin-orbit interaction (blue and red dashed lines correspond to the minority and majority spin, respectively). However, the sizes of the resulting gaps are different for both directions. 

By comparison of Fig. 3 (c) and Fig. 3 (e) we deduce that the difference between the $\alpha^+$ and $\alpha^-$ sheets observed in the experiment (Fig. 3 (c)) clearly resemble changes expected in the bulk electronic structure for the directions  $ k\parallel\vec{M}$ and $k\perp\vec{\text{M}}$ (Fig. 3 (e)). Therefore, we have shown that the experimentally observed effect is of the same character and within the same order of magnitude as expected for the SOI-related shifts of the bulk electronic bands linked to the change of the remanent magnetization direction. However, the exact shape of the experimental Fermi sheet found for the magnetization RIGHT (Fig. 3 (c)), \textit{i.e.}, lack of the symmetry with respect to the $\overline{\Gamma}$ - $\overline{\text{X}}$ line cannot be explained solely by the bulk electronic structure. 
In order to shed more light onto this issue we need to address the role that the surface plays in our photoemission experiment. Figure 4 shows the result of the slab calculations which reveal the surface electronic structure of Fe(001) for the magnetization lying in plane (Fig. 4(a)) and out of plane (Fig. 4 (b)). Blue (red) symbols correspond to the predominantly minority (majority) states.

\begin{figure*}
	\includegraphics{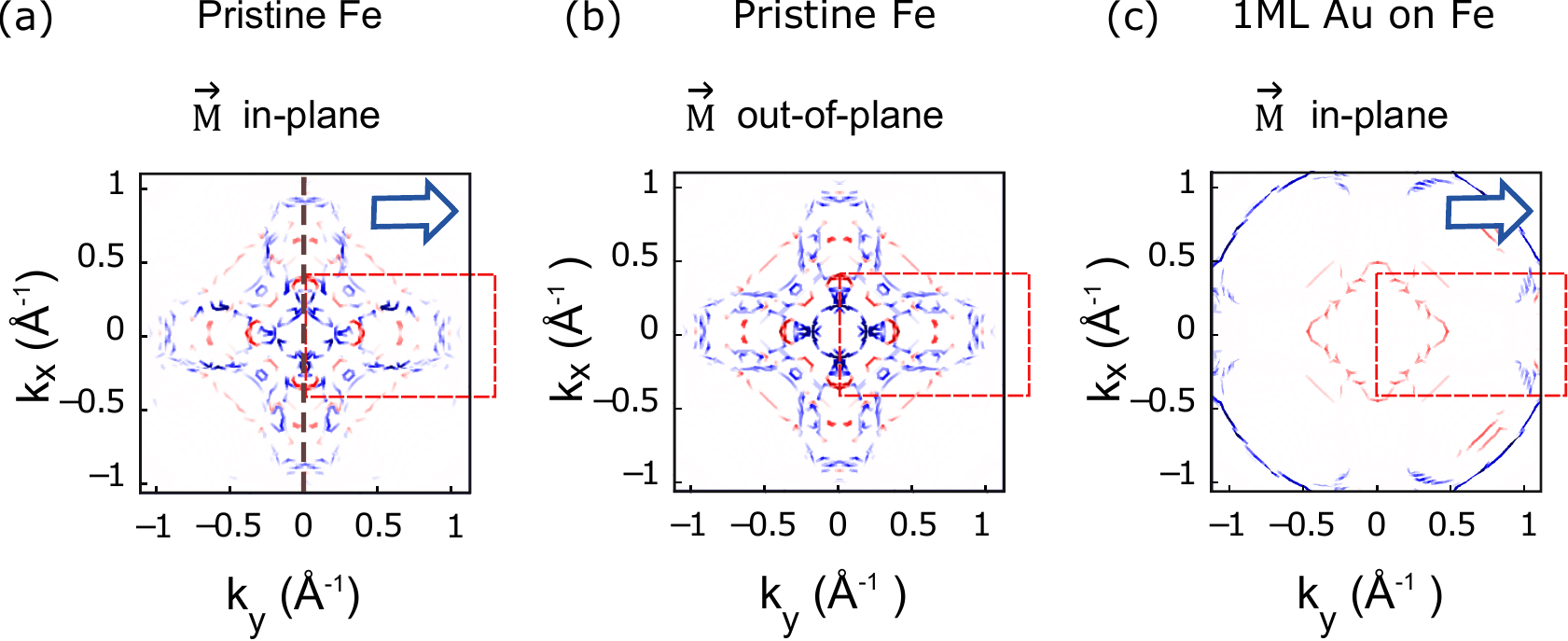}
	\caption{
		Results of the relativistic GGA slab calculations that reveal the surface electronic structure of Fe(001) (a) Pristine Fe(001), magnetization in-plane. Dashed vertical line marks the mirror plane. (b) Pristine Fe(001), magnetization out-of-plane. (c) 1 ML Au/Fe(001), magnetization in-plane. Blue (red) symbols correspond to the predominantly minority (majority) states. The red rectangles show the fraction of the \textit{k}-space measured in the experiment (Fig. 3b). \label{Fig4}}
\end{figure*}

The first question we need to address is why the experimentally observed electronic band structure resembles so well the bulk dispersions (Fig. 3 (a) and (b)), and does not reflect the theoretically predicted surface electronic structure (Fig. 4). The calculations reveal a well-known surface state (SS) \cite{Chantis2007} that forms a four-petal-like shape at the Fermi surface. The signatures of it were observed experimentally in the past \cite{Turner1984,Turner1983,Plucinski2009, Vescovo1993,Sawada1999}. One of the possible reasons why the SS does not show up in our experiment might be the photon-energy-dependent experimental photoionization cross section of this state. Another possibility is that the SS is quenched by Au atoms, as they tend to diffuse towards the surface of Fe(001), even though a cryogenic deposition temperature (T = 50 K) was used in order to suppress this effect. To test this hypothesis, we performed additional GGA slab calculations, with 1ML of Au placed onto the surface of Fe(001). The result is presented in Fig. 4 (c). We see that the presence of Au atoms shifts this minority Fe surface state above the Fermi level, and thus makes it inaccessible in a photoemission experiment (see Fig. 6 in the APPENDIX D for the calculated band dispersions). We also observe that additional states are introduced by the presence of Au, but mostly for the binding energies close to $E_B$ = 2 eV, which is below the energy range studied in our experiment. Even though the surface states predicted by the calculation are not revealed in our experiment, the results presented in Figs. 4 (a) and (b) reflect the influence of the magnetization direction onto the symmetry of the electronic states when the translational symmetry of the system is broken by the presence of the surface. When we consider a magnetic case of the Fe(001) surface, its symmetry is $C_4$ when magnetized out-of-plane (90$^\circ$ rotation axis) (Fig. 4 (b)) and only $C_s$ (single mirror plane, no rotational symmetry) if magnetized in plane (Fig. 4 (a)). In such a situation of symmetry breaking by the surface and in-plane magnetization, electronic bands observed for two opposite \textit{k}-vectors that are perpendicular to the magnetization direction (namely, +$k_x$ and -$k_x$) are not equivalent. This effect can be seen within the area marked by the dashed red rectangle in Fig. 4 (a). The very same type of asymmetry was found in our experiment for the magnetization RIGHT (Fig. 3 (c)). 
In order to confirm that the symmetry of the states measured with ARPES is consistent with the symmetry expected for the surface states, we measured ARPES spectra around the same $\overline{\text{X}}$ point with respect to the experimental reference frame for three other remanent magnetization directions, \textit{i.e.}, LEFT, UP and DOWN. The obtained Fermi surfaces are summarized in Figs. 5 (a)-(d). Pronounced differences in the electronic structure depending on the magnetization direction are clearly recognized. These magnetization-dependent electronic features were perfectly reproducible for multiple remagnetization cycles.

Figure 5 (a) presents the result of the measurement for the magnetization RIGHT, which was already shown in Fig. 3. Now we can compare it to the spectral weights observed for other magnetization directions. We see that the gap found for the $\alpha^+$ band for the magnetization RIGHT (Fig. 5 (a)) switches to the band $\alpha^-$  when the magnetization is pointing LEFT (Fig. 5 (b)). On the contrary, when the magnetization direction points UP or DOWN (Fig. 5 (c) and (d), respectively) both $\alpha^+$ and $\alpha^-$ are rather symmetric with respect to the $\overline{\Gamma}$ - $\overline{\text{X}}$ ($k_x = 0\, \text{\AA}^{-1}$) line. However, the difference observed between these two magnetization directions is the shift of both $\alpha^+$ and $\alpha^-$  sheets towards higher $k_y$ values for the magnetization DOWN, as compared to UP. The revealed shifts are of the order of $0.1\, \text{\AA}^{-1}$. The illustration presented in Fig. 5 (e) shows how the measured spectra can be arranged with respect to the magnetization direction when the role of the light incidence direction is neglected, which allows to identify the symmetries characteristic for the observed bands. The superimposed black solid line in the illustration shows the theoretical cut through the Fermi surface of bcc Fe at $k_{\perp} \sim 2.2\, \text{\AA}^{-1}$. The experimental result reveals only single mirror plane (vertical dashed line), perpendicular to the magnetization direction, \textit{i.e.}, the symmetry point group $C_s$, which is identical to the one found for the electronic structure of the surface (Fig. 4 (a)). Therefore, we confirmed that the experimentally observed electronic bands show the symmetry expected for the surface states. Next, we check if this experimental observation is reflected in the one-step model calculations of the photoemission process.

\begin{figure*}
	\includegraphics{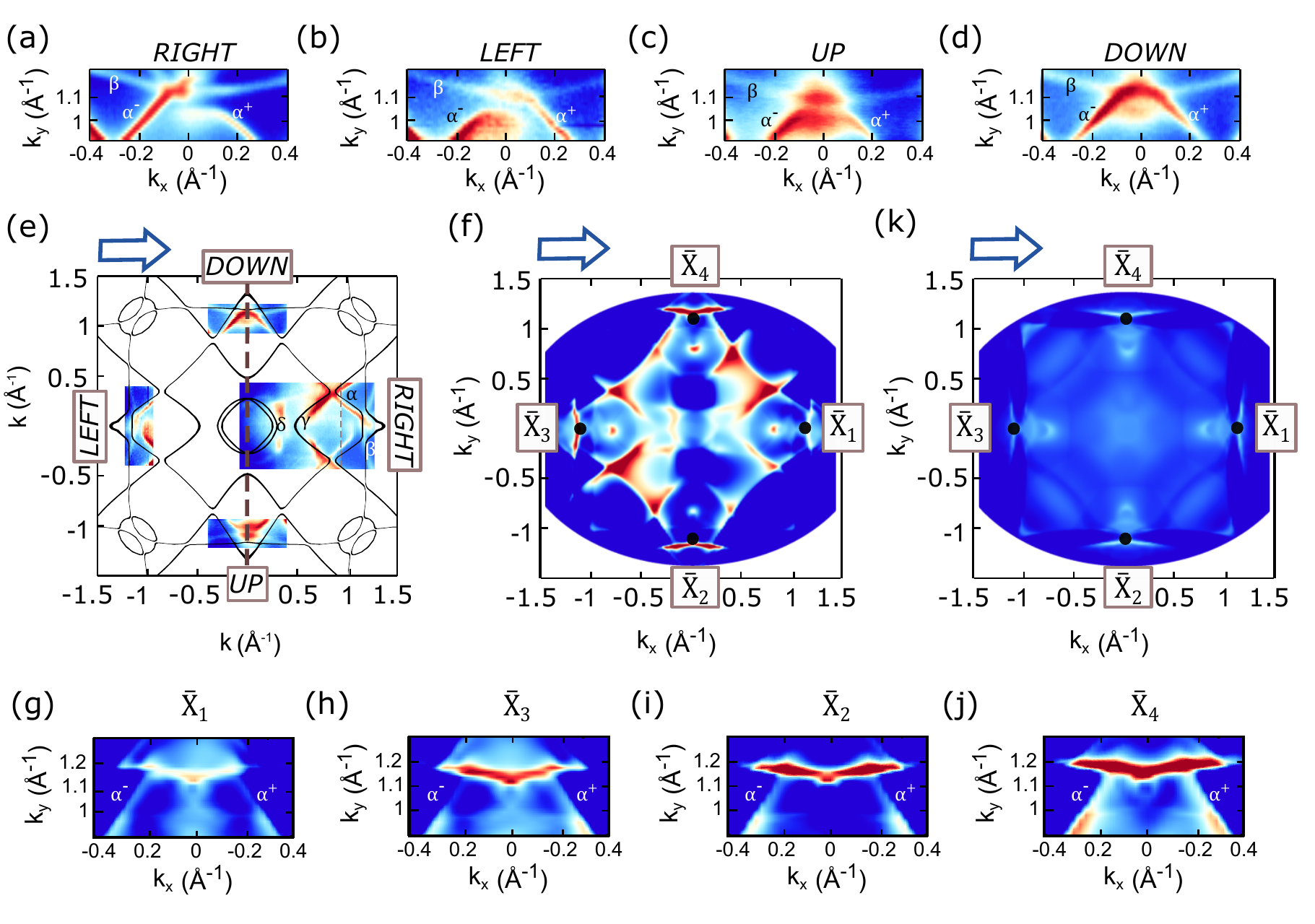}
	\caption{
		(a-d) Electronic structure of Fe(001) close to the $\overline{\text{X}}$ point of the SBZ for four different in-plane easy magnetization directions measured at $h\nu$ = 16.8 eV. (e) Fermi surfaces derived from (a-d) arranged with respect to the fixed magnetization direction (arrow); this approximation is applicable when the role of the light incidence direction is neglected. The black line depicts the electronic band structure of the bulk $k_{\perp}$ = 2.2$\,\text{\AA}^{-1}$ calculated using the \textit{GW} method. Dashed vertical line marks the mirror plane. (f) Result of the one-step model photoemission calculation that shows spectral intensities within the entire SBZ. The arrow marks the magnetization direction. (g-j) Results of the one-step model photoemission calculation shown for the fractions of the \textit{k}-space that correspond to experimental Fermi surfaces presented in (a-d). (k) Result of the one-step model photoemission calculations performed without the \textit{d}-\textit{f} transitions. All the one-step model photoemission calculation results shown are obtained with intentionally quenched surface contribution and the light incidence direction along the magnetization.  \label{Fig5}}
\end{figure*}

 The result of the one-step model calculations of photoemission induced by light of $h\nu$ = 16.8 eV from Fe(001) in the geometry of our experiment is presented as a Fermi surface in Fig. 5 (f). Figures 5 (g-j) show magnified regions of the one-step model calculation results in the fractions of the \textit{k}-space that correspond to the measurements shown in Fig. 5 (a-d). The projection of the light incidence direction on the sample surface lies along the magnetization direction (indicated by an arrow in Fig. 5 (f)). In the calculation, the spectral intensity originating from the surface states was intentionally quenched, according to the very good correspondence of our experiment and the bulk electronic structure of bcc Fe. In this way, we can theoretically examine how the bulk states located close to the sample surface are affected by the broken translational symmetry. Immediately obvious is the close resemblance of the results of the one-step model calculations, the results of the experiment and the theoretical bulk electronic structure cut at $k_{\perp} \sim 2.2\, \text{\AA}^{-1}$ (Fig. 5 (e) and (f)). This means that the results of the one-step model calculations are consistent with the assumptions of the free-electron final state model. Very good agreement concerning the intensity of the photoemission features between experiment and one-step model calculations is revealed, especially for the $\gamma$ Fermi sheet (Fig. 5 (e) and (f)). The \textit{k}-space region near the $\overline{\text{X}}$ points is also relatively well reproduced by the calculations. The comparison of Fig. 5 (a-d) and (g-j) shows that even though the exact shifts of the $\alpha$ sheet observed in experiment are not visible in the calculations, the experimental band symmetries are well reproduced. The \textit{k}-space regions in the vicinity of the points $\overline{\text{X}}_1$ (Fig. 5 (g)) and $\overline{\text{X}}_3$ (Fig. 5 (h)) (that correspond to the measurements performed with the magnetization pointing RIGHT (Fig. 5 (a)) and LEFT (Fig. 5 (b)), respectively) are not symmetric with respect to the $k_x$ = 0 line. The sheets $\alpha^+$ are less (more) intense than $\alpha^-$ for the point $\overline{\text{X}}_1$ ($\overline{\text{X}}_3$), which resembles the experimentally found shift of the $\alpha$ sheet. Both $\alpha$ branches are less intense near the $\overline{\text{X}}_2$  point (Fig. 5 (i)) than near the point $\overline{\text{X}}_4$ (Fig. 5 (j)), which corresponds to the bigger distance between the $\alpha$ and $\beta$ Fermi sheets observed in the experiment for the magnetization UP (Fig. 5 (c)). 

In addition, to access the information on the final states involved in the observed transitions we performed one-step model photoemission calculations neglecting the excitations from \textit{d} to \textit{f} states. The result is presented in Fig. 5 (k). We see that the remaining spectral intensity is relatively weak, and the spectral features that we could compare to the experimental result vanished. Therefore, we showed that the electronic transitions seen in our ARPES experiment are mainly from the initial states of \textit{d} character to the final states of \textit{f} character.
  
Based on the extensive analysis of the experimental results based on the calculations we can conclude that we experimentally observed the spin-orbit interaction-related and magnetization-dependent band gaps within the bulk states of bcc Fe. The asymmetry of the bulk electronic bands found in our experiment is consistent with the one expected for a system where both exchange coupling and spin-orbit interaction coexist, when the existence of the sample surface is taken into account. 

Even though Fe belongs to the model ferromagnets and the electronic structure of Fe has been intensively studied by experiments and theory in the last decades \cite{Turner1983,Turner1984,Plucinski2009,Vescovo1993,Sawada1999,Ackermann1984,Heimann1978}, to the best of our knowledge, the only spectroscopic proof of a modification of the electronic band structure as a response to a change of the magnetization direction was provided for Fe(110) thin films in the scanning tunneling spectroscopy study of Bode \textit{et al.} \cite{Bode2002}. Up to now, \textit{k}-resolved experimental observations of the magnetization-related changes of the electronic band structure of a ferromagnetic film were reported only for surface- or interface- related electronic states such as surface state of oxidized Gd(0001) \cite{Krupin2005} or quantum well states in Co/W(110) \cite {Moras2015}. In these works, the observed effect was interpreted in relation to the so-called Rashba term \cite{Bychkov1984,Bihlmayer2006} in the total Hamiltonian of the system. In a theoretical study of the TAMR effect in Fe(001) \cite{Chantis2007} the resonant surface bands (not observed in our experiment) were shown to depend on the magnetization direction which was also attributed to the Rashba effect. Rashba interaction occurs when the electron moves in the system that lacks structural inversion symmetry (all surfaces and interfaces), in the presence of the electric field and spin-orbit interaction \cite{Bihlmayer2006}. Because in the reference frame of the electron an electric field is seen as an effective magnetic field, a shift of the electron energies occurs. Rashba interaction of the electron with spin and momentum $\hbar \vec{k}$ within the electric field directed along the \textit{z}-axis ($\vec{e_z}$) is expected, according to: $\Delta E=\alpha_R(\vec{e_z}\times \vec{k}) $ where $\alpha_R$ represents the Rashba parameter, which depends on the strength of the electric field and spin-orbit interaction \cite{Bihlmayer2006}. In case of ferromagnetic materials, for the magnetization along $k_y$, the Rashba shift is expected to be positive for $k_x >$ 0 and negative for $k_x <$ 0 \cite{Krupin2005}. Therefore, from the symmetry point of view, it resembles the situation discussed for the surface electronic structure of Fe(001) (Fig. 4), and observed in our experiment. In principle, the interface between Fe(001) and Au(001) could be a source of high spin-orbit interaction which might enhance the Rashba effect in Fe \cite{Park2013}. Such a proximity effect might become observable near the sample surface due to the small amount of Au which is probably present at the Fe surface. In principle, the Rashba interaction affects electronic states that are localized near surfaces or interfaces; the typical examples are (i) two dimensional electron gas and (ii) surface states of the high-\textit{Z} crystalline materials. In our experiment we clearly observed that magnetization direction modifies the bulk electronic states of Fe(001). Due to the surface sensitivity of ARPES (equal to few monolayers), the bulk electronic bands visible in our experiment were probed in the near-surface region. Rashba-type effects for bulk bands were observed before for the occupied bulk bands of Bi(111) \cite{Kimura2010}, and unoccupied bulk bands of Au(111) \cite{Wissing2013} and explained as a result of the reflection of the Bloch states from the surface \cite{Krasovskii2011}. Therefore, we speculate that the observed magnetization-dependent changes in the electronic band structure are related to the Rashba-type effect acting on the near-surface part of the bulk-like electronic states.

\section{IV. SUMMARY AND CONCLUSIONS}

The manipulation of the Fermi surface using an external magnetic field was demonstrated for a prototypical ferromagnet, Fe(001). We have shown the comprehensive ARPES data with clear modifications of the bulk electronic bands of Fe(001) in response to the remanent change of the magnetization direction. 
The experimental electronic band structure of Fe(001) was found to match well the results of the \textit{GW} first-principles theoretical calculations for bulk bcc Fe combined with the assumptions of the free electron final state model, which allowed to identify the experimentally observed electronic bands. Results of the one-step model photoemission calculations reproduce symmetries observed in the experiment for different magnetization directions which indicates that the observed bulk electronic bands are modified by the proximity of the surface. The observed symmetries reveal the interplay between exchange-coupling and spin-orbit coupling in the experimental configuration where surface and in-plane remanent magnetization contribute to the symmetry breaking. The observed symmetries agree with the ones expected for the Rashba-type effect.  
ARPES spectra were found to bear fingerprints of the distinct magnetization direction, which means that a new way of determining the in-plane magnetization direction based on the photoemission spectra was identified.
We interpret the observed effect as the result of the opening of spin-orbit interaction- and magnetization-related band gaps, existence of which is essential for the emergence of the fundamental magnetic phenomena like magnetocrystalline anisotropy and x-ray magnetic linear dichroism. What is more, the detected electronic band gaps might play a substantial role in the spin transport effects. Our finding shows that, contrary to the common belief, spin-orbit coupling cannot be neglected in the analysis of the electronic properties even for elements as light as \textit{3d} ferromagnets.

\begin{acknowledgments}
\section{ACKNOWLEDGEMENTS}	
	
	This work was supported by the Helmholtz Association and Alexander von Humboldt Foundation. We thank B. K\"upper and A. Bremen for the technical assistance. HE, JM, JB and SB thank the BMBF (05K13WMA), the DFG (FOR 1346), and the COST Action MP
	1306 for financial support. JM also thank the CENTEM PLUS (LO1402), CENTEM (CZ.1.05/2.100/03.0088) for financial assistance.
	
\end{acknowledgments}

\appendix
\section{Appendix A: DETAILS OF THE \textit{GW} CALCULATIONS}
For the starting point of \textit{GW} we used the LDA exchange-correlation functional \cite{Perdew1981} and the code FLEUR \cite{FLAPW}. For this, we used an angular momentum cutoff of $l_{max}$=8 in the atomic spheres and a plane-wave cutoff of 5.0 bohr$^{-1}$ in the interstitial region. We used the experimental lattice parameter 2.87 $\text{\AA}$ from Ref. \cite{Wyckoff1963} and the semicore 3\textit{s}- and 3\textit{p}-states of Fe were treated as valence orbitals by the use of local orbitals. The mixed product basis \cite{Friedrich2010,Kotani2002} used in the \textit{GW} calculation was constructed with an angular momentum cutoff of $l_{max}$=4 and a plane-wave cutoff of 3.0 bohr$^{-1}$. We used 170 unoccupied bands and a 10$\times$10$\times$10 \textit{k}-point sampling of the Brillouin zone. This leads to energy differences and a Fermi energy converged up to 10 meV. Two additional local orbitals per angular momentum up to $l$=3 were included to describe high-lying states accurately and to avoid linearization errors \cite{Friedrich2006,Friedrich2011}.
\section{Appendix B: DETAILS OF THE ONE STEP MODEL CALCULATIONS}
For the calculation of the ground-state properties we used a fully relativistic multiple scattering method in the framework of density functional theory (Korringa-Kohn-Rostoker, KKR) \cite{Ebert2012}. We used the tight-binding (\textit{TB}) approximation for an effective calculation of the surface properties. This method provides a fast convergence of the calculation of the so-called \textit{TB} structure constants \cite{Zeller1995}. The calculations of the self-consistent potentials have been carried out in the atomic sphere approximation. For the exchange-correlation functional we used the parameterization of Vosko, Wilk and Nusair \cite{Vosko1980}.  

Using this method we are able to investigate the influence of the surface contribution to the resulting ARPES spectra. This is done via tuning the multiple scattering between surface barrier and 2D semi-infite system as follows. The bulk potentials are well ordered according to the two-dimensional crystal structure of the given surface. This means that the spin-orbit interaction determined from the single ion core potentials which contain no information about $k_{\perp}$ is transformed into a $k_{\parallel}$-splitting of the surface state through multiple scattering. This scattering procedure is realized by the bulk reflection matrix \textit{B} which is developed into the two-dimensional reciprocal lattice vectors of the corresponding surface. This matrix represents the complete electronic structure information of the semi-infinite bulk. The scattering is in first order proportional to \textit{B}. When the off-diagonal elements of \textit{B} are reduced the transfer of spin-orbit interaction from the bulk to the surface layer is diminished. For more details the reader is referred to \cite{Nuber2011}. 

\section{Appendix C: DETAILS OF THE FREE ELECTRON FINAL STATE MODEL}
According to the free electron final state model, excitation with a fixed photon energy leads to photoemission from well-defined regions of the BBZ. Following the formula that relates the perpendicular component of the electron wave vector ($k_{\perp}$) with the kinetic energy of the photoemitted electrons ($E_{kin}$), the emission angle ($\Theta$) and so-called inner potential V$_0$ (adjusted to be equal to 7eV):$k_{\perp}=\sqrt{(2m/\hbar^2 E_{kin} \cos^2(\Theta)+V_0)} $ we estimate that using $h\nu$=16.8 eV corresponds to performing a cut of the BBZ zone at $k_{\perp} \sim 2.2 \text{\AA}^{-1}$.
\section{Appendix D: SURFACE ELECTRONIC STRUCTURE OF 1ML Au/Fe(001)}
The band dispersions corresponding to the Fermi surfaces presented in Fig. 4 of the main text are presented below (Fig. 6). The cut along the high symmetry direction ($\overline{\Gamma}$ - $\overline{\text{X}}$) is shown for the pristine Fe(001) surface and for 1~ML Au/Fe(001) (Fig. 6(a) and Fig. 6(b), respectively). The magnetization lies in plane, along the $k_y$-axis. Blue (red) color mark the bands of the predominantly minority (majority) character. The size of the marker corresponds to the localization of the state within the surface layer. The asymmetry between the positive and negative $k_x$ direction introduced by the presence of the surface is clearly visible, especially for the 1~ML Au/Fe(001) (Fig. 6 (b)). 

\begin{figure}
	\includegraphics[width=8cm]{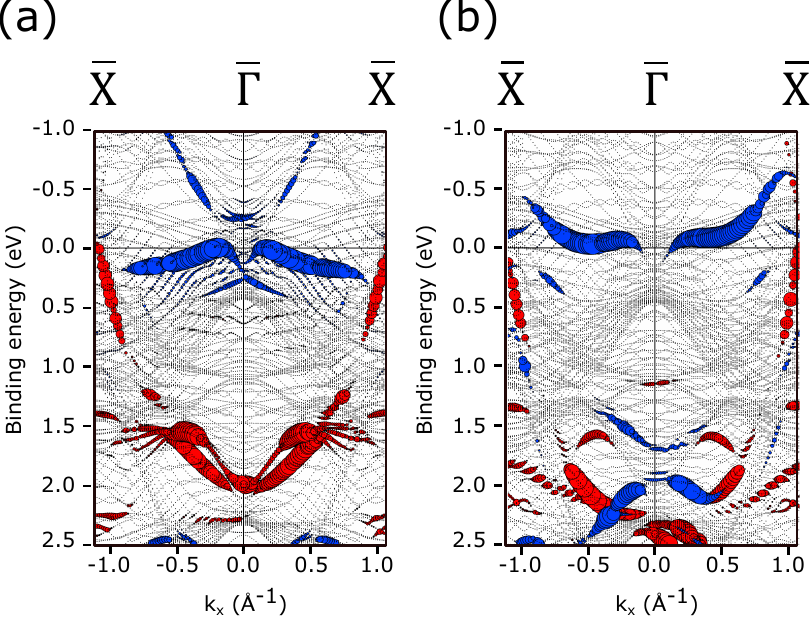}  
	\caption{
		Results of the GGA relativistic slab calculations showing surface electronic structure of Fe(001) and 1~ML Au/Fe(001). Dispersions along $\overline{\Gamma}$ - $\overline{\text{X}}$ direction for pristine Fe surface (a) and 1~MLAu/Fe(001) (b);    Blue (red) symbols mark bands with predominantly minority (majority) spin character. Size of the symbol corresponds to the localization in the surface layer. Magnetization lies in plane, along the $k_y$-axis.
	}	\label{Fig_suppl}
\end{figure}


%

\end{document}